\UseRawInputEncoding
 \documentclass[a4paper,twocolumn,preprintnumbers,citeautoscript,aps,prd,longbibliography,superscriptaddress,floatfix,nofootinbib]{revtex4}

\usepackage{amsmath,amsfonts,amssymb,graphics}
\usepackage[normalem]{ulem}
\usepackage{comment}
\usepackage{url}
\usepackage{graphicx}
\usepackage{cancel}

\usepackage{mathrsfs}
\usepackage{bbm}
\usepackage{pifont}
\usepackage{braket}
\usepackage{siunitx}
\usepackage{slashed}

\usepackage{mathtools}
\usepackage{xspace}
\usepackage{booktabs}
\usepackage{csquotes}
\usepackage[dvipsnames,svgnames]{xcolor}
\usepackage{tikz}
\usetikzlibrary{positioning,shapes,arrows}
\usetikzlibrary{decorations.pathmorphing}
\usetikzlibrary{decorations.markings}
\usepackage{standalone}

\usepackage{url}
\usepackage[bookmarks=false, breaklinks]{hyperref} 

\definecolor{nicered}{rgb}{0.5,.0,.0}
\definecolor{darkblue}{rgb}{0,.1,.9}
\definecolor{lightblue}{rgb}{0,.1,.6}
\definecolor{applegreen}{rgb}{0.55, 0.71, 0.0}
\definecolor{darkgreen}{rgb}{0.0, 0.2, 0.13}
 
\hypersetup{colorlinks, citecolor=darkgreen ,linkcolor=nicered, urlcolor= lightblue}

\newcommand{\gl}[1]{{\color{black}#1}}

\newcommand{\Sp}{{\rm Sp}}
\newcommand{\SU}{{\rm SU}}

\begin{document}

\title{Axion-Like Electrophilic Portal for Pion Dark Matter}

\author{Vincenzo Fiorentino}
\email{vincenzo.fiorentino@lnf.infn.it}

\affiliation{Istituto Nazionale di Fisica Nucleare, Laboratori Nazionali di Frascati, C.P. 13, 00044 Frascati, Italy}
\affiliation{Dipartimento di Fisica, Universit\`a Roma Tre, Via della Vasca Navale 84, I-00146 Rome, Italy}

\author{Ji-Heng Guo}
\email{guojiheng@buaa.edu.cn}

\affiliation{Istituto Nazionale di Fisica Nucleare, Laboratori Nazionali di Frascati, C.P. 13, 00044 Frascati, Italy}
\affiliation{School of Physics, Beihang University, Beijing 102206, China}

\author{Giacomo Landini}
\email{glandini@lnf.infn.it}

\affiliation{Istituto Nazionale di Fisica Nucleare, Laboratori Nazionali di Frascati, C.P. 13, 00044 Frascati, Italy}

\author{Federico Mescia}
\email{federico.mescia@lnf.infn.it}

\affiliation{Istituto Nazionale di Fisica Nucleare, Laboratori Nazionali di Frascati, C.P. 13, 00044 Frascati, Italy}

\begin{abstract}

We investigate a scenario where Strongly Interacting Massive Particle (SIMP) dark matter interacts with an axion-like particle (ALP) that couples exclusively to electrons. This minimal setup provides interactions which enforce thermal equilibrium between dark matter and the SM in the early Universe. We analyze the cosmological evolution of the dark sector and the constraints arising from dark matter annihilations, ALP laboratory searches and astrophysical observations. Our results show that the allowed  parameter space is wider than previous studies and
an ALP with mass $m_a \sim {\cal O}(10)~\text{MeV}$ can act as a viable portal between the visible and dark sectors. Interestingly, this mass range overlaps with the parameter space suggested by the reported $X_{17}$ anomaly. 
Furthermore, the introduction of non-vanishing $\theta$ angle in the dark sector of the model opens up the parameter space to heavy ALP masses.
\end{abstract}

\twocolumngrid
\maketitle

\section{Introduction}

Strongly Interacting Massive Particles (SIMPs) have emerged as a compelling class of dark-matter candidates~\cite{Hochberg:2014kqa,Hochberg:2015vrg, Kuflik:2015isi,Bernal:2015bla, Bernal:2015xba, Bernal:2015ova, Choi:2015bya, Choi:2016hid,Soni:2016gzf, Kamada:2016ois,  Kamada:2017tsq,Bernal:2017mqb,Cline:2017tka, Choi:2017mkk,  Kuflik:2017iqs, Heikinheimo:2018esa, Choi:2018iit,Hochberg:2018rjs, Bernal:2019uqr, Choi:2019zeb,  Katz:2020ywn, Smirnov:2020zwf, Xing:2021pkb, Braat:2023fhn,Garani:2021zrr, Dey:2016qgf, Bernreuther:2023kcg,Chu:2024rrv,Chu:2025hga,Garcia-Cely:2024ivo,Garcia-Cely:2025flv,Davighi:2024zip,Davighi:2025awm}, offering an alternative to more traditional weakly interacting scenarios. SIMPs arise in models where the dark matter production is driven by 3-to-2  annihilations and the cosmological relic abundance is reproduced for dark matter masses in the MeV-GeV range. 
This class of models has attracted significant interest because it generically predicts an elastic self--interaction cross section per unit mass of order $\mathcal{O}(\text{cm}^2/\text{g})$. Such values can help reconcile the cuspy central densities predicted by $N$--body simulations of collisionless cold dark matter~\cite{Dubinski:1991bm, Navarro:1995iw, Navarro:1996gj} with a range of astrophysical observations that instead favor shallower, core-like profiles~\cite{Dave:2000ar, Vogelsberger:2012ku, Rocha:2012jg, Peter:2012jh, Elbert:2014bma, Fry:2015rta} (see also~\cite{Tulin:2017ara} for a review on self-interacting dark matter).
A natural realization of the SIMP setup arises in models of dark pions. These particles emerge as pseudo-Goldstone bosons of a confining dark sector analogous to QCD, in which the so-called Wess-Zumino-Witten (WZW) term provides the required 3-to-2 annihilations.

A crucial aspect of SIMP models is the interaction between dark matter and Standard Model (SM) particles. The dark matter production mechanism relies on the assumption that the dark sector remains in thermal equilibrium with the SM bath in the early Universe at the same temperature. Such equilibrium can be maintained through the presence of a mediator that enables interactions between the dark and visible sectors.
One possibility is that the mediator is a dark photon (see, for example,~\cite{Hochberg:2015vrg,Garcia-Cely:2025flv,Katz:2020ywn}).
Another option is to consider a pseudo-scalar portal, as in~\cite{Kamada:2017tsq,Hochberg:2018rjs}.

In this work, we reconsider the scenario in which a pseudo-scalar particle mediates the interaction between dark pions and the Standard Model. {This mediator may arise as the pseudo-Goldstone boson of a spontaneously broken global U(1) symmetry,} and we generically denote it as an Axion-Like Particle (ALP).

We focus specifically on scenarios where the ALP couples exclusively to electrons~\cite{Eberhart:2025lyu} (see also~\cite{Darme:2020sjf,Altmannshofer:2022ckw,Ferreira:2022xlw,Adhikary:2026rck,Arias-Aragon:2025kiz})\footnote{The coupling of ALPs to charged leptons has been studied also in~\cite{Bauer:2017ris,Cornella:2019uxs,Calibbi:2020jvd,Bauer:2021mvw,Bertuzzo:2022fcm,Armando:2023zwz,Ema:2025bww,Ferreira:2025qui,Hostert:2025ffy,Altmannshofer:2026opc}.}. By contrast,  previous studies~\cite{Kamada:2017tsq,Hochberg:2018rjs} considered ALP-photon interactions.
Restricting the ALP to couple only to electrons relaxes several constraints associated with tree-level ALP-photon couplings and opens new regions of viable parameter space. We first examine the regime of light ALPs and show that the allowed parameter space is substantially larger than previously thought, accommodating ALP masses down to $\sim \mathcal{O}(10)$ MeV.

This possibility is particularly intriguing because, within this mass range, the ALP could be connected to the tentative $17~\text{MeV}$ resonance reported by the PADME experiment~\cite{PADME:2025dla}.
Although the overall significance reaches only about $2\sigma$, the result remains noteworthy, as it aligns, within current uncertainties, with the longstanding ATOMKI anomalies~\cite{Krasznahorkay:2015iga,Krasznahorkay:2018ip,Krasznahorkay:2019apb,Krasznahorkay:2019njf,Krasznahorkay:2021joi,Krasznahorkay:2022prc,Krasznahorkay:2023x17,Krasznahorkay:2024universe} observed in nuclear transitions, which have long hinted at a possible new boson, commonly denoted $X_{17}$
~\cite{Arias-Aragon:2025wdt,Barducci:2025hpg,Abdallah:2024uby,Alves:2023ree,Denton:2023gat,Barducci:2022lqd,Feng:2020mbt,Zhang:2020ukq,Feng:2016jff,Feng:2016ysn}.

Next, we study the impact of a non-vanishing topological $\theta$ angle in the dark sector of our model. 
Such a parameter is generically allowed in QCD-like dark sectors and, unlike in the SM, there is no experimental motivation to set it to zero. 
We show that a non-zero $\theta$ angle induces additional interactions that offer an alternative mechanism for establishing thermal equilibrium with the SM bath. 
This selects a different region of the parameter space, permitting ALPs heavier than the dark matter.

The outline of the paper is the following: we review  the dark matter model in Section~\ref{sec:SIMP}. In section~\ref{sec:ALP} we introduce the electrophilic  ALP portal and discuss the conditions for dark matter thermalization and the relevant constraints. We also mention the possibility to identify the ALP with a 17 MeV resonance. In Section~\ref{sec:theta} we examine the consequences of a non-vanishing $\theta$ angle on the available parameter space.  
We present our conclusion in Section~\ref{sec:conclusions}. We briefly comment on the prospects of an ALP portal in connection with the dark-matter model of Ref.~\cite{Garcia-Cely:2024ivo} in Appendix~\ref{app:resonant}.

\section{The dark matter model}\label{sec:SIMP}

We consider the dark matter (DM) model  introduced in Ref.~\cite{Hochberg:2018rjs}, which consists of a QCD-like confining sector based on an $\Sp(2N_c)$ gauge group, with $N_c \geq 2$. The matter content includes $2N_f$ Weyl fermions $q_i$ (with $i = 1,\dots, 2N_f$) transforming in the fundamental representation of the gauge group. The corresponding Lagrangian is given by
\begin{equation}\label{eq:LagSIMP}
    \mathcal{L}_{\rm DM}=-\frac{1}{4}G_{\mu\nu}^2+q^\dagger i\bar{\sigma}^\mu D_\mu q-\frac{1}{2}(e^{i\theta}M_{ij}q_iq_j+{\rm h.c.})\,.
\end{equation}
Here $G_{\mu\nu}$ is the $\Sp(2N_c)$ gauge field strength tensor and $M$ is the fermion mass matrix. We assume a degenerate mass spectrum $M_{ij}=m\,(\gamma_{2N_f})_{ij}$ where $m>0$ and $(\gamma_{2N_f})_{ij}$ is the $\Sp(2N_f)$ invariant tensor\footnote{The invariant tensor of $\Sp(2N_f)$ is defined as $\gamma_{2N_f}=\mathbb{I}_{N_f}\otimes i\sigma_2$. Under a generic $\Sp(2N_f)$ transformation denoted by $V$, it satisfies $V^T\gamma_{2N_f}V=\gamma_{2N_f}$, leaving the mass term in Eq.~\eqref{eq:LagSIMP} invariant.}. 

We include in the mass matrix a physical phase $\theta$. A chiral rotation of the fermion fields, $q_i \to e^{-i\theta/2} q_i$, removes this phase from the mass term. However, since the transformation is anomalous under the $\Sp(2N_c)$ gauge interactions, it generates a topological $\theta\,G_{\mu\nu}\widetilde{G}^{\mu\nu} $ term in the Lagrangian\footnote{Notice that, with the normalization of Eq.~\eqref{eq:LagSIMP}, the topological term is normalized as $(g^2/32\pi^2)N_f\theta\, G_{\mu\nu}\widetilde{G}^{\mu\nu}$, where $g$ is the gauge coupling.}.

In the massless limit $m=0$, the Lagrangian in Eq.~\eqref{eq:LagSIMP} is invariant under a global $\SU(2N_f)$ symmetry, corresponding to independent rotations of each fermion field $q_i\to e^{i\alpha_i}q_i$.
In analogy with ordinary QCD, we expect that gauge interactions confine at some energy scale $\Lambda$, giving rise to a fermion condensate $\langle{q_iq_j\rangle}\propto(\gamma_{2N_f})_{ij}$, which spontaneously breaks the global symmetry group of Eq.~\eqref{eq:LagSIMP} as $\SU(2N_f)\to \Sp(2N_f)$. This gives rise to $N_\pi=N_f(2N_f-1)-1$ pseudo-Goldstone bosons $\pi^a$, from now on referred to as dark pions. The fermion mass matrix chosen in Eq.~\eqref{eq:LagSIMP} preserves the $\Sp(2N_f)$ symmetry and generates a potential for the pseudo-Goldstone bosons. Their interactions are described by Chiral Perturbation Theory (ChPT) as (for a review of standard ChPT techniques see, e.g~\cite{Pich:1995bw,Scherer:2002tk})
\begin{equation}\label{eq:chiLag1}
   \mathcal{L}_\pi= \frac{f_\pi^2}{16}{\rm Tr}[\partial_\mu U^\dagger\partial^\mu U]-\frac{f_\pi^2}{8}B_0{\rm Tr}[e^{i\theta}MU+{\rm h.c.}]\\
   +\mathcal{L}_{\rm WZW}\,,
\end{equation}
where we introduced the meson field
\begin{equation}
     U=e^{2i\pi/f_\pi}\,(\gamma_{2N_f})\,, \qquad \pi=\pi^a\lambda^a\,.
\end{equation}
Here $f_\pi$ is the dark pion decay constant, $B_0$ is related to the fermion condensate as $\langle{qq\rangle}\sim B_0f_\pi^2$ and  $\Lambda\sim 2\pi f_\pi$. The matrices $\lambda^a$ belong to the coset\footnote{The generators belonging to the coset $\SU(2N_f)/\Sp(2N_f)$ are explicitly given by
\begin{equation*}
    \frac{1}{\sqrt{2}}T_{\rm asym}\otimes\sigma_k,\qquad \frac{1}{\sqrt{2}}T_{\rm sym}\otimes \mathbb{I}_2\,,
\end{equation*} 
where $\{T_{\rm asym},T_{\rm sym}\}$ are the antisymmetric and symmetric generators of $\SU(N_f)$, respectively. For $N_f=2$ we have $T_{\rm asym}=\sigma_2$ and $T_{\rm sym}=\{\sigma_1,\sigma_3\}$.
}
$\SU(2N_f)/\Sp(2N_f)$ corresponding to the (spontaneously) broken generators, normalized as ${\rm Tr}[\lambda^a\lambda^b]=2\delta^{ab}$, so that the mass of the dark pions is given by $m_\pi^2=2B_0 m \cos \theta$. The dark pions are the lightest states of the model and the unbroken flavor $\Sp(2N_f)$ guarantees their stability, so that they are a natural candidate for dark matter.

The Wess--Zumino--Witten (WZW) term
\begin{equation}
    \mathcal{L}_{\rm WZW}
    = \frac{8N_c}{15\pi^2 f_\pi^5}\,
      \epsilon^{\mu\nu\rho\sigma}\,
      \epsilon_{abcde}\,
      \pi^a \partial_\mu \pi^b
      \partial_\nu \pi^c
      \partial_\rho \pi^d
      \partial_\sigma \pi^e\,,
\end{equation}
induces the $3\pi \to 2\pi$ annihilation processes that govern the thermal evolution of the dark sector. As long as these interactions remain efficient, the dark pions follow their equilibrium distribution. This holds until the Universe cools to the freeze-out temperature $T_{\rm fo} \sim m_\pi/20$, when the $3 \to 2$ reaction rate drops below the Hubble expansion rate and the dark pion abundance freezes out.

More precisely, the relic density is obtained by solving the Boltzmann equation
\begin{equation}
    \dot{n}_\pi + 3H n_\pi
    = -\left( n_\pi^3 - n_\pi^2 n_\pi^{\rm eq} \right)
      \langle \sigma v^2 \rangle_{3\pi \to 2\pi}\,,
\end{equation}
where $n_\pi$ is the dark pion number density, $n_\pi^{\rm eq}$ its equilibrium value, 
$    H=\sqrt{4\pi^3g_\rho/45}\,T^2/M_{\rm Pl}$
is the Hubble expansion rate (with $g_\rho$ the number of relativistic degrees of freedom in the thermal bath, $g_\rho \sim \mathcal{O}(10)$ within the SM at $T \sim \mathrm{MeV}-\mathrm{GeV}$), and $\langle \sigma v^2 \rangle_{3\pi \to 2\pi}$ is the thermally averaged cross section for the WZW-induced $3\pi \to 2\pi$ processes. The equilibrium density is given by 
$n_\pi^{\rm eq}=(N_\pi m_\pi^2 T/2\pi^2)K_2(m_\pi/T)$.

We solved the Boltzmann equation numerically for several values of $N_c$. The observed dark matter relic abundance is obtained for dark pion masses in the range
\begin{equation}\label{eq:range}
 m_\pi \sim 100~\mathrm{MeV} - 1~\mathrm{GeV},   
\end{equation}
and $m_\pi \lesssim 2\pi f_\pi$, in agreement with Ref.~\cite{Hochberg:2014kqa}. The exact value of $f_\pi$ is fixed by solving the Boltzmann Equation for each choice of dark matter mass.
\gl{The DM particles undergo elastic scatterings with cross section per unit mass of order $\mathcal{O}(\text{cm}^2/\text{g})$. Constraints derived from observational data from galaxy clusters~\cite{Harvey:2015hha,Bondarenko:2017rfu,Harvey:2018uwf,Sagunski:2020spe,Eckert:2022qia,DES:2023bzs} can be satisfied in the range of DM masses in Eq.~\eqref{eq:range} for appropriate values of $N_c$\footnote{\gl{Notice that, values of DM masses $250\text{ MeV}\lesssim m_\pi\lesssim 700$ MeV are allowed for any number of colors $N_c\geq2$. Larger masses, up to $m_\pi\sim 1$ GeV, requires $N_c\gtrsim 4$. For lower masses, a larger number of colors is required to avoid cluster constraints on self-interactions.}}.}

\section{The electrophilic ALP portal}\label{sec:ALP}

The Boltzmann equation described above relies on the crucial assumption that the dark pions  remain in thermal equilibrium with the SM bath at least until DM freeze-out, such that both sectors share the same temperature $T$. This requires the presence of an explicit portal between the dark sector and the SM. In its absence,  number changing processes within the dark sector would modify the evolution of the dark sector temperature and the corresponding dark pion population.

The portal is not unique and many options have been investigated: Ref.~\cite{Hochberg:2015vrg} studied a dark photon portal (see also~\cite{Katz:2020ywn,Garcia-Cely:2025flv}), while Refs.~\cite{Kamada:2017tsq,Hochberg:2018rjs} discuss a pseudo-scalar particle $a$ coupled to dark pions and photons\footnote{In Refs.~\cite{Davighi:2024zip,Davighi:2025awm}, a topological portal between the dark and the visible sectors was considered.}. 
More specifically, in Ref.~\cite{Kamada:2017tsq} $a$ must have roughly the same mass as the dark pion, $m_a\simeq m_\pi$, so that the DM cosmological evolution is dominated by semi-annihilations $\pi \pi\to \pi a$ rather than $3\pi\to2\pi$ processes. This assumption is relaxed in Ref.~\cite{Hochberg:2018rjs}
where $a$ keeps the dark sector and the SM in thermal equilibrium as long as $100\text{ MeV}\lesssim m_a<m_\pi$.

In this paper, we reconsider the pseudo-scalar portal scenario, under the assumption that the pseudo-scalar particle is electrophilic, i.e., it only couples to electrons~\cite{Eberhart:2025lyu}\footnote{A leptophilic pseudo-scalar portal has been considered in Ref.~\cite{Armando:2023zwz} in the case of fermionic dark matter. A pseudo-scalar portal for scalar dark matter has also been investigated in~\cite{DEramo:2025xef}.}.

\subsection{Coupling to electrons }

We consider the electrophilic interaction given by\footnote{As emphasized in~\cite{Eberhart:2025lyu}, two distinct electrophilic models can be considered, depending on whether the interaction term in Eq.~\eqref{eq:LagALP} is pseudo-scalar $a\, \bar{e}\gamma_5 e$ or derivative $\partial_\mu a\, \bar{e}\gamma^\mu \gamma_5 e$. In the derivative scenario, a large coupling to photons is induced at one-loop when $m_a \gg m_e$. Here we concentrate on the ``pseudo-scalar model'' of Ref.~\cite{Eberhart:2025lyu}.}
\begin{equation}\label{eq:LagALP}
\mathcal{L}_{a}=\frac{1}{2}\partial_\mu a\partial^\mu a-\frac{1}{2}m_a^2a^2-g_{ae}a\bar{e}i\gamma_5e\,.
\end{equation}
{The pseudo-scalar particle $a$ can arise as the pseudo-Goldstone boson of a global U(1)$_\text{GB}$ symmetry spontaneously broken at some UV scale $f_a$. These particles are usually referred to as ALPs.  A well-known example is the QCD axion~\cite{Wilczek:1977pj,Weinberg:1977ma,Peccei:1977hh}, motivated by the Strong CP problem. 

A UV-complete model { involving extra scalars charged under U(1)$_\text{GB}$} can generate  the interaction in Eq.~\eqref{eq:LagALP}  from a Yukawa coupling, which, upon spontaneous symmetry breaking, gives rise to

\begin{equation}
    -m_e \bar{e}_Le_R e^{iC_ea/f_a}+\text{h.c.}\,
\end{equation}
Expanding the exponential function, we reproduce Eq.~\eqref{eq:LagALP}. Then, 
the relation between the electron coupling and the U(1)$_\text{GB}$ breaking scale is given by  $C_e/f_a=g_{ae}/m_e$, where in typical UV completions $C_e\sim\mathcal{O}(1)$. 
The mass term $m_a$ is introduced as a soft explicit breaking of the U(1)$_\text{GB}$ symmetry and its origin depends on the UV completion of the theory.\footnote{A plausible possibility is that quantum-gravity effects induce Planck-suppressed U(1)-breaking operators that generate the mass of the ALP.
}

In Eq.~\eqref{eq:LagALP} there is no UV contribution\footnote{A UV contribution to the ALP-photon coupling would come by loops of heavy fermions, electrically charged and chiral under the global U(1)$_\text{GB}$ symmetry~\cite{Armando:2023zwz}. In the absence of such fermions, no UV contribution is generated.} to the ALP-photon coupling $a F_{\mu\nu}\widetilde{F}^{\mu\nu}$. This is generated at one loop by a triangle diagram involving the electron interaction of Eq.~\eqref{eq:LagALP}, but is strongly suppressed when $m_a \gg m_e$, allowing the model to evade several experimental bounds~\cite{Eberhart:2025lyu}.

Moreover, performing a field redefinition, the Lagrangian in Eq.~\eqref{eq:LagALP} can be equivalently  written as 
\begin{equation}\label{eq:LagALPder}
\begin{split}
\mathcal{L}_{a}&=\frac{1}{2}\partial_\mu a\partial^\mu a-\frac{1}{2}m_a^2a^2\\ &+C_e\frac{\partial_\mu a}{2f_a}\bar{e}\gamma^\mu \gamma_5e+\frac{C_{e}}{f_a}\frac{\alpha_{\rm em}}{4\pi}aF_{\mu\nu}\widetilde{F}^{\mu\nu}\,.
\end{split}
\end{equation}
Now, the invariance of $a$ under shift  symmetry (broken by the anomalous term) is manifest. As noticed in Ref.~\cite{Eberhart:2025lyu},  Eq.~\eqref{eq:LagALPder} generates the same effective coupling with 
photons\footnote{The tree-level term of $aF_{\mu\nu}\widetilde{F}^{\mu\nu}$ is canceled by the one-loop contribution induced by the derivative interaction.} of Eq.~\eqref{eq:LagALP}.
In the following, we adopt the basis of Eq.~\eqref{eq:LagALP} to perform our computations.
}

\vspace{0.2cm}

The ALP decays into electron-positron pairs $a\to e^+e^-$ with rate
\begin{equation}\label{eq:decayrate}
    \Gamma_a=\frac{g_{ae}^2m_a}{8\pi}\sqrt{1-\frac{4\,m_e^2}{m_a^2}}\,,
\end{equation}
{while the decay rate to photons is negligible.}
We assume $m_a>\mathcal{O}(1)$ MeV, so that its lifetime is much shorter than $1$ sec for $g_{ae}\gtrsim 10^{-10}$, thus evading all cosmological constraints. We remain agnostic about the UV completion, although possible realizations have been discussed in Ref.~\cite{Eberhart:2025lyu}. 

\gl{We comment that the choice of electrophilic interactions is mostly due to phenomenological reasons. Universal ALP-lepton couplings could also be considered as a viable option.  In such a case, the constraints on the dark sector and its interactions with the SM would be largely unaffected, whereas laboratory ALP searches would impose stronger constraints. The model would  still be viable, but the allowed parameter space would be more restricted.}

\subsection{Coupling to the Dark Sector}

The ALP couples to the dark fermions as

\begin{equation}\label{eq:aqLag}
    \mathcal{L}_{aq}=-\frac{1}{2}(e^{i(\theta+a/f_a)}M_{ij}q_iq_j+{\rm h.c.})\,.
\end{equation}

This operator can arise, for example, from a Yukawa interaction with a scalar field $\Phi\sim f_ae^{ia/f_a}$, if both $\Phi$ and the dark fermions $q_i$ are charged under  U(1)$_\text{GB}$.  In such a case, a small Yukawa coupling is needed to justify the hierarchy $m\ll f_a$. 

More generally, a concrete UV model involving extra fields charged under U(1)$_\text{GB}$, similar in spirit to the typical UV completions for the QCD axion~\cite{DiLuzio:2020wdo}, can generate an anomalous coupling of the ALP to the $\Sp(2N_c)$ gauge bosons, $(g^2/32\pi^2)N_f aG_{\mu\nu}\widetilde{G}^{\mu\nu}/f_a$, see also~\cite{Kamada:2017tsq}\footnote{In general, there is no reason why the full ALP potential, determined by UV physics, is aligned with the potential generated by the $G_{\mu\nu}\widetilde{G}^{\mu\nu}$ term. Thus, the value of the ALP field in the minimum of its potential is not expected to cancel the $\theta$ phase. We define the ALP field so that the minimum of its potential corresponds to $a=0$.}. Then, at energies below $f_a$, an anomalous fermion field redefinition, $q_i\to e^{ia/2f_a}q_i$, removes the $aG_{\mu\nu}\widetilde{G}^{\mu\nu}$ term and provides the interaction in Eq.~\eqref{eq:aqLag}, while keeping $f_a$ and $m$ as independent parameters.

Starting from Eq.~\eqref{eq:aqLag}, we can derive  the effective Lagrangian describing the interactions of the ALP with the dark pions by simply replacing $M\to e^{ia/f_a}M$ in Eq.~\eqref{eq:chiLag1}.  Notice that for the consistency of the effective Lagrangian description we require $f_a\gtrsim 2\pi f_\pi$ (and thus $f_a\gtrsim m_\pi)$. In the rest of this section we assume $\theta=0$, while we will relax this assumption in Sec.~\ref{sec:theta}.

As first observed in Ref.~\cite{Hochberg:2018rjs}, we specialize in the case $N_f=2$ so that ${\rm Tr}[\pi^3]=0$ (as well as ${\rm Tr}[\pi^5]=0$)  and the DM relic abundance is controlled by $3\pi\to2\pi$ processes induced by the WZW term\footnote{For a generic choice of gauge group and number of flavors $N_f$, the interaction $a {\rm Tr}[\pi^3]$ is present and gives rise to semi-annihilations $\pi\pi\to\pi a$.}. The relevant interaction term in the Chiral Lagrangian is

\begin{equation}\label{eq:alpPionLag}
    \frac{m_\pi^2}{8f_a^2}a^2{\rm Tr}[\pi^2]=\frac{m_\pi^2}{4f_a^2}a^2\pi^a\pi^a\,,
\end{equation}

which gives rise to elastic scatterings $\pi a\to\pi a$ as well as annihilations $\pi\pi\to aa$ if $m_\pi>m_a$.
Finally, interactions with the dark pions generate a contribution to the ALP mass, namely $\Delta m_a=2m_\pi^2f_\pi^2/f_a^2$. In order to avoid fine-tuned cancellations we require that $m_a^2>\Delta m_a^2$.

\vspace{0.2cm}
Up to this point, we have discussed the scenario in which $a$ is the pseudo-Goldstone boson of a U(1) symmetry. In the following, we will also consider the option where $a$ is a generic pseudo-scalar particle, for which the coupling $g_{ae}$ and the scale $f_a$ are unrelated and can therefore be treated as independent parameters. From now on, we will refer to the pseudo-scalar particle as an ALP, independently of its pseudo-Goldstone nature. When it is the case, we will specify whether the pseudo-Goldstone interpretation is viable.

\subsection{Dark Matter thermalization and constraints}

\begin{figure*}[t]
\!\!\!\includegraphics[width=0.45\textwidth]{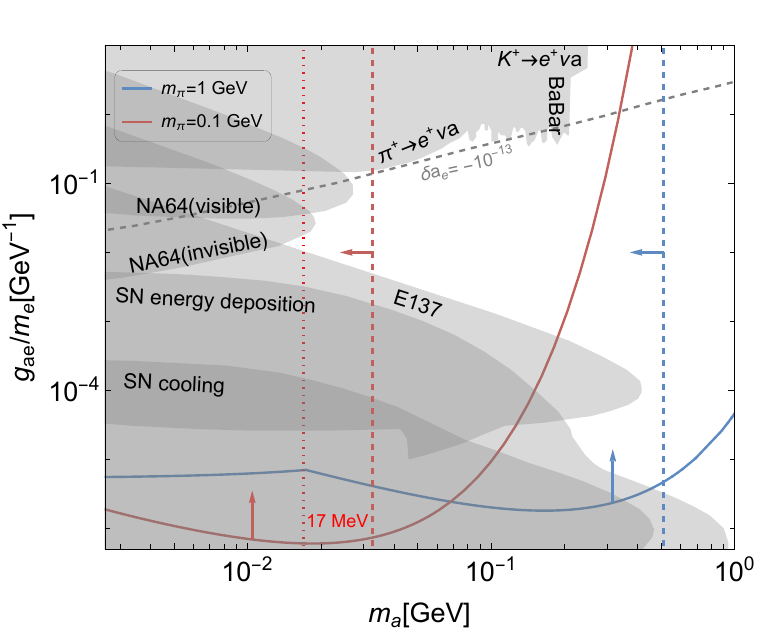}
\includegraphics[width=0.45\textwidth]{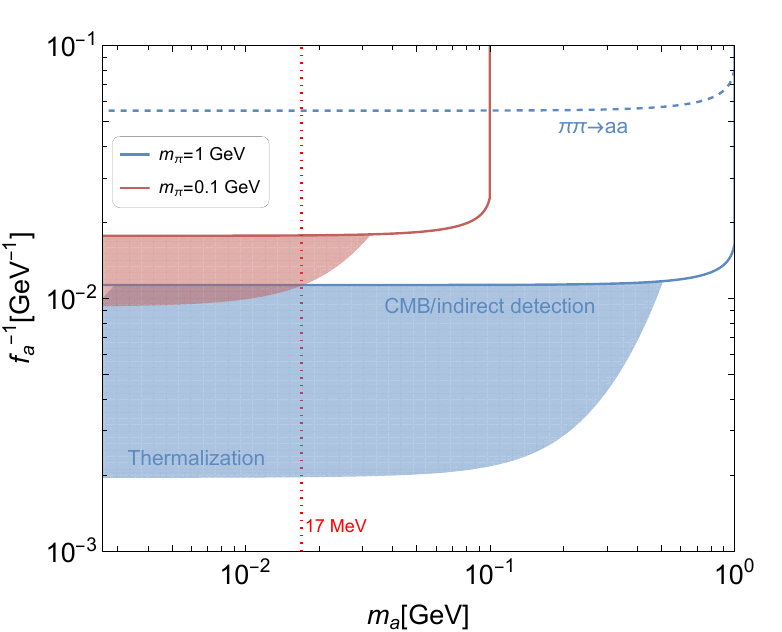} \,\,\,\,\,\,
\caption{{ Summary of the constraints for different choices of dark matter mass. Left: thermalization between the ALPs and the SM is efficient above the solid lines. The maximal value of $m_a$ allowed from CMB and indirect detection constraints is shown as a dashed line. The gray region is excluded from  laboratory searches and SN1987A. The available parameter space corresponds to the white region above the solid line and at the left of the dashed line, as shown by the arrows. }   { Right: the available parameter space corresponds to the colored regions. These are bounded from above by CMB and indirect detection constraints and from below by the ALP-dark matter thermalization condition.  The red dotted line corresponds to an ALP of mass $m_a=17$ \text{MeV}. }
}
 \label{fig:alp1}
 \end{figure*}

The interactions mediated by the ALP portal must be sufficiently strong to maintain thermal equilibrium between the dark pions and the SM particles for as long as the $3\pi\to2\pi$  processes remain active.

This is achieved in two steps:

\begin{enumerate}
\item[(\textit{i})] elastic scatterings $\pi a\to \pi a$ transfer energy from the dark  pions to the ALP sector and keep them in kinetic equilibrium at a common temperature, while 
\item[(\textit{ii})] the ALP thermalizes with the SM bath through decays $a\to e^+e^-$ and scatterings $a\,e^\pm\to \gamma\,e^\pm$, together with their corresponding inverse processes.
\end{enumerate} When (\textit{i}) and (\textit{ii}) are simultaneously satisfied, the dark pions, the ALP and the SM bath share the same temperature $T$. This must hold until (at least) dark matter freeze-out.

The condition (\textit{i}) is independent of the specific interactions between the ALP and the SM sector and has been studied in~\cite{Hochberg:2018rjs}. For the reader's convenience, we briefly review it. Given the interaction in Eq.~\eqref{eq:alpPionLag}, the dark pions elastically scatter off ALPs with cross section 
\begin{equation}
    \sigma v(\pi a\to \pi a)=\frac{m_\pi^4 \sqrt{m_a^4-2m_a^2(m_\pi^2+s)+(m_\pi^2-s)^2}}{8 \pi f_a^4 (s^2-(m_\pi^2-m_a^2)^2)}\,,
\end{equation}
where $v$ is the relative velocity and the sum (average) over the degrees of freedom of the final (initial) states has been performed. To keep the dark pions and the ALP in kinetic equilibrium at a common temperature, the elastic scatterings have to be fast enough. To ensure this, one needs to compare: (\textit{a}) the rate at which energy is transferred  through the elastic scatterings, which tend to establish a common temperature, (\textit{b}) the rate of $3\pi\to2\pi$ processes, which increase the pion temperature by converting mass into kinetic energy, and (\textit{c}) the Hubble expansion rate. By
comparing these rates, we  find, in agreement with~Ref.~\cite{Hochberg:2018rjs}, that kinetic equilibrium is maintained until the temperature drops below $T_{\rm dec}\simeq m_\pi(\sqrt{5\pi g_\rho}f_a^4/m_\pi^3 M_{\rm Pl})^{1/4}$ if $m_a\ll m_\pi$. If the ALP is non-relativistic during DM freeze-out, $m_a\gtrsim m_\pi/20$, its number density gets suppressed by the Boltzmann factor $n_a\propto e^{-m_a/T}$, so that the rate of elastic scatterings drops exponentially and thermalization is much harder. The precise condition is obtained by solving the Boltzmann equation for the energy transfer upon computation of the elastic collision rate, see the discussion in Appendix A of~\cite{Hochberg:2018rjs}. To understand the dependence on the parameters, we observe that close to the freeze-out temperature the rate of $3\pi\to2\pi$  processes is of the order of  $\simeq H$. Thus, kinetic equilibrium is established if $(T/m_\pi)n_a^{\rm eq}\langle{\sigma v\rangle}_{\pi a}\gtrsim H$. We require that $T_{\rm dec}\leq T_{\rm fo}$, so that kinetic equilibrium is kept until dark matter freeze-out.

On the other hand, Eq.~\eqref{eq:alpPionLag} also gives rise to $\pi\pi\to aa$ annihilations. In the non-relativistic limit, they proceed with thermally averaged cross section 

\begin{equation}
    \langle{\sigma v\rangle}_{\pi\pi}\equiv\langle{\sigma v\rangle}(\pi\pi\to aa) \simeq  \frac{m_\pi^2}{64\pi N_{\pi} f_a^4}\sqrt{1-\frac{m_a^2}{m_\pi^2}}\,.
\end{equation}
These processes must be sub-leading compared to the $3\pi\to2\pi$ annihilations, to not modify the computation of the relic abundance. We simply require that the corresponding interaction rate is smaller than the Hubble rate at the time of dark matter freeze-out, $n_\pi\langle{\sigma v}\rangle_{\pi\pi}\leq H|_{T=T_{\rm fo}}$. 

Much more stringent constraints arise from measurements of the Cosmic Microwave Background (CMB) and indirect detection searches. Indeed, dark matter annihilations $\pi\pi\to aa$ are followed by ALP decays $a\to e^+e^-$. 
On the one hand, the electrons and positrons produced by dark matter annihilations can scatter off the photons of the CMB and modify the evolution of recombination, leaving an imprint on the CMB anisotropies. On the other hand, X-ray telescopes can detect  the photons produced via Inverse Compton scattering of background radiation off electrons originating from DM annihilations. Since ALP decays are practically instantaneous and proceed with branching ratio equal to one, the constraints apply directly\footnote{All the bounds are obtained for annihilations into $e^+e^-$. Since every annihilation process $\pi\pi\to aa$ gives rise to a $e^+e^-e^+e^-$ final state, a small correction could be expected.} on $\langle{\sigma v\rangle}_{\pi\pi}$. CMB measurements set the strongest bounds~\cite{Cirelli:2024kph,Cirelli:2024ssz,Slatyer:2015jla} $\langle{\sigma v\rangle}_{\pi\pi}\lesssim 5\times 10^{-29}\, \text{cm}^3/\text{sec}$ for dark matter masses below $\sim 200$ MeV.
For heavier DM masses, up to $\sim 5$ GeV, the strongest constraints are provided by X-ray telescopes such as XMM-Newton telescope~\cite{Koechler:2023ual}, $\langle{\sigma v\rangle}_{\pi\pi}\lesssim10^{-28}\, \text{cm}^3/\text{sec}$. Notice that these constraints are slightly stronger with respect to those used in~\cite{Hochberg:2018rjs}, where the ALP decays to photons.

We show the parameter space in the right plot of Fig.~\ref{fig:alp1}, as a function of the scale $f_a$, for two representative values of the dark matter mass $m_\pi$. For each choice of $m_\pi$, the colored area represents the allowed region, which lies around $f_a\sim\mathcal{O}(100)$ GeV. Smaller values of $f_a$ lead to stronger dark matter annihilations, which are excluded by CMB and indirect detection constraints. Notice that dark matter annihilations to ALPs dominate over $3\pi\to2\pi$ annihilations only in regions which are already excluded (above the dashed lines, only visible for $m_\pi=1$ GeV in the plot). On the other hand, for larger values of $f_a$, dark matter scatterings off ALPs are not efficient enough to establish kinetic equilibrium. Finally, we {notice that a different choice of $m_\pi$, in the range of Eq.~\eqref{eq:range}, would lead to an allowed region which interpolates between those shown in Fig.~\ref{fig:alp1}.}

\subsection{ALP-electron thermalization}

In this subsection we address the condition (\textit{ii}). The relevant processes are ALP decays $a\to e^+e^-$ and scattering processes $a\, e\to \gamma\, e$, together with their corresponding inverse reactions. 

The ALP decays with thermally averaged decay rate $\langle{\Gamma_a\rangle}=(\,K_1(m_a/T)/K_2(m_a/T)\,)\Gamma_a$. The
inverse decays $e^+e^-\to a$ proceed with rate $\langle{\Gamma_{\rm ID}\rangle}=(n_a^{\rm eq}/n_e^{\rm eq})\langle\Gamma_a\rangle$, which is exponentially suppressed at temperatures $T<m_a$. Such rates should be faster than the Hubble expansion at the time of dark matter freeze-out. This leads to

\begin{equation}\label{eq:therm1}
    \frac{\Gamma_a m_a}{4 H T}e^{-m_a/2T}\bigg|_{T=T_{\rm fo}}\gtrsim1\,,
\end{equation}
see the Appendix~\ref{app:alpelectron} for more details.

At the same time, the ALPs scatter off  electrons and positrons converting to photons with thermally averaged cross section

\begin{equation}\label{eq:aecross}
\langle{\sigma v(ae^\pm\to\gamma e^\pm)\rangle}=\frac{\gamma_{ae}^{\rm eq}}{n_a^{\rm eq}n_e^{\rm eq}}\,,
\end{equation}
in terms of the interaction rate density
\begin{equation}
   \gamma_{ae}^{\rm eq}=\frac{T}{32\pi^4}\int_{(m_a+m_e)^2}^\infty ds \,\sqrt{s}\,K_1(\frac{\sqrt{s}}{T})\int dt \frac{\overline{|\mathcal{M}_{ae}|^2}}{8\pi s}\,,
\end{equation}
where in the amplitude we perform the average over the initial state (denoted by the bar) and the sum over the final state.
In the limit in which the electron mass can be neglected, the amplitude of the process is given by 

\begin{equation}
\overline{\left|\mathcal{M}_{ae}\right|^2}=8\pi\alpha_{\rm em}\, g^2_{ae}
\frac{t^2+m_a^4}{s\left(s+t-m_a^2\right)}\,,
\end{equation}
where $s$ and $t$ are the usual Mandelstam variables.
We require that the scattering rate is faster than the Hubble rate
\begin{equation}\label{eq:therm2}
    \Gamma_{\rm scatt}=\sum_{f=\{e^+,e^-\}}n_f^{\rm eq}\langle{\sigma v(af\to\gamma f)\rangle}\gtrsim H\,,
\end{equation}
at the time of dark matter freeze-out.
In our numerical calculations we adopt a more precise condition, obtained by solving the Boltzmann equation for the energy transfer rate, see details in Appendix~\ref{app:alpelectron}. 
The ALPs are kept in equilibrium with the SM bath as long as at least one of the conditions in Eq.~\eqref{eq:therm1} and Eq.~\eqref{eq:therm2} is satisfied. Decays and inverse decays are more efficient if $m_a\gtrsim T_{\rm fo}$, while scattering processes are more efficient for very light ALPs, $m_a\ll T_{\rm fo}$. Indeed, when the ALP is relativistic, $T\gg m_a$, its decay rate is suppressed by the boost factor~$\sim m_a/T$, while the scattering rate is proportional to the number density $\propto T^3$.

\begin{figure*}[t]
\includegraphics[width=0.45\textwidth]{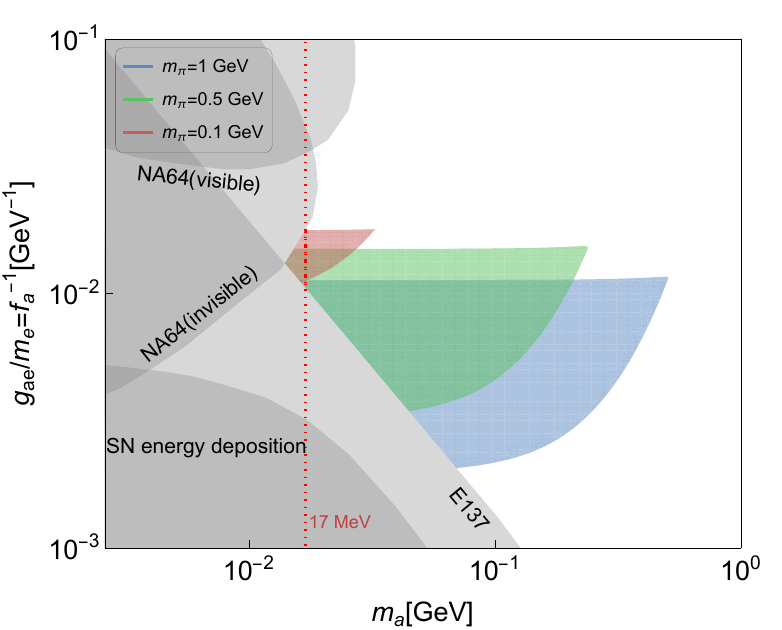} \,\,\,\,\,\,
\caption{We show for the pseudo-Goldstone interpretation of the ALP, $g_{ae}/m_e\equiv 1/f_a$,  the combination of all the constraints of Fig.~\ref{fig:alp1}. The red  (green) [blue]  area is allowed for $m_\pi=0.1(0.5)[1]$ GeV. The gray area is excluded by laboratory and SN constraints, while the white region is excluded by CMB/indirect detection and failure to achieve an efficient ALP-DM thermalization, see the caption of Fig.~\ref{fig:alp1} and the main text for details. The value of $g_{ae}$ is always sufficiently large to achieve an efficient ALP-SM thermalization. The red dotted line corresponds to $m_a=17$ MeV.}
 \label{fig:alp3}
 \end{figure*}

\subsection{ALP constraints}

In the parameter space of our interest,
electrophilic ALPs are mostly constrained by laboratory searches, while  cosmological constraints are not relevant~\cite{Eberhart:2025lyu}. Indeed, ALPs heavier than $\mathcal{O}(2.5)$ MeV do not contribute to the effective number of neutrinos $N_{\rm eff}$. Furthermore, as already pointed out, the lifetime of our ALP is much shorter than 1 sec, thus avoiding constraints from Big Bang Nucleosynthesis.
On the other hand, some constraints on electrophilic ALPs come from the cooling and explosion  of the Supernova SN1987A. However, these affect only low values of $g_{ae}/m_e\lesssim 10^{-3}\text{ GeV}^{-1}$ and are not competitive with laboratory searches (see, for instance~\cite{Fiorillo:2025sln,Ferreira:2022xlw}). 

 The strongest constraints are set by beam-dump experiments such as E137~\cite{Bjorken:1988as} performed at SLAC and NA64~\cite{NA64:2019auh} (both in the visible and invisible modes) at CERN. The ALP particle is produced from the interaction of the beam with the target mostly via Dark Bremsstrahlung and Primakoff processes and subsequently decays  inside (visible) or outside (invisible) the detector. Large couplings, $g_{ae}/m_e\gtrsim0.1\text{ GeV}^{-1}$, are also constrained by the BaBar~\cite{BaBar:2014zli} collaboration, reinterpreting the constraints for dark photon production and decay\footnote{Notice that for an electrophilic ALP, the bounds are re-interpreted up to the muon production threshold $m_a<2m_\mu\simeq 200$ MeV~\cite{Eberhart:2025lyu}.} ($e^+e^-\to \gamma a$, followed by $a\to e^+e^-)$ and by searches of rare meson decays $\pi^+\to e^+\nu a$~\cite{SINDRUM:1986klz} and $K^+\to e^+\nu a$~\cite{Poblaguev:2002ug}. {Finally, the anomalous magnetic moment of the electron $a_e=(g_e-2)/2$ does not impose any binding constraint to an electrophilic ALP,  see the discussion in Ref.~\cite{Eberhart:2025lyu}.}

\vspace{0.3cm}

We show the available parameter space in the left plot of Fig.~\ref{fig:alp1}, in terms of the ALP-electron coupling $g_{ae}/m_e$. The regions shaded in gray are excluded by laboratory searches for electrophilic ALPs and SN1987A. The solid lines correspond to the ALP-electron thermalization condition, namely the weaker of Eq.~\eqref{eq:therm1} and Eq.~\eqref{eq:therm2}, for different values of dark matter mass $m_\pi$. Thermalization is efficient for values of the coupling above the lines, as indicated by the arrows. In most of the parameter space, thermal equilibrium is kept through ALP decays, while scatterings become more relevant only at low $m_a$, thus explaining the kink around $m_a\sim m_\pi/50$.  The dashed lines represent the maximal value of $m_a$ allowed by CMB and indirect detection constraints (see the plot on the right).

In conclusion, for each value of $m_\pi$, the allowed parameter space corresponds to the white region above the solid line and on the left of the dashed line. On the one hand, as the dark matter mass grows, the allowed region gets wider. On the other hand, large values of $m_a$ are excluded. Indeed, whenever the ALP becomes heavier than the dark matter, its number density --- and thus the relevant rates for thermalization --- drops exponentially and thermalization is not achieved. We conclude that the electrophilic ALP portal works for $m_a\lesssim m_\pi$.  Compared to previous studies, the parameter space is enlarged at low values of the ALP mass. Indeed, masses as low as $\mathcal{O}(10)$ MeV are allowed in our scenario for each choice of DM mass. This is due to the fact that the electrophilic scenario evades the most stringent constraints arising from the ALP-photon coupling. In particular, we highlight that an ALP with mass of 17 MeV can be accommodated in our setup, as shown by the red dotted lines in both plots of Fig.~\ref{fig:alp1}. 

Finally, we note that the allowed region of parameter space is consistent with interpreting the ALP as a pseudo--Goldstone boson of a global $U(1)$ symmetry broken at the scale $f_a$. Indeed, by combining the two panels of Fig.~\ref{fig:alp1}, we find that there exist regions, in particular around $m_a = 17~\text{MeV}$, in which all constraints can be simultaneously satisfied while preserving the relation $g_{ae}/m_e \sim 1/f_a$. This is shown explicitly in Fig.~\ref{fig:alp3}, where we impose the condition $g_{ae}/m_e = 1/f_a$.

\subsection{Hints of an $e^+ e^-$ excess at 17 MeV}

Recent PADME data~\cite{PADME:2025dla} hint at a possible excess in $e^{+}e^{-}$ annihilation around a 
center-of-mass energy of $\sqrt{s} \approx 16.9~\text{MeV}$. Although the global significance 
is only at the $\sim 2\sigma$ level, the observation is intriguing because it is consistent, 
within uncertainties, with the earlier ATOMKI tensions~\cite{Krasznahorkay:2015iga,Krasznahorkay:2018ip,Krasznahorkay:2019apb,Krasznahorkay:2019njf,Krasznahorkay:2021joi,Krasznahorkay:2022prc,Krasznahorkay:2023x17,Krasznahorkay:2024universe} observed in nuclear transitions. 
This can be potentially accommodated by introducing a new boson commonly referred to as $X_{17}$~\cite{Arias-Aragon:2025wdt,Barducci:2025hpg,Abdallah:2024uby,Alves:2023ree,Denton:2023gat,Barducci:2022lqd,Feng:2020mbt,Zhang:2020ukq,Feng:2016jff,Feng:2016ysn}.
Nevertheless, the broader experimental picture remains 
highly dynamic, especially in light of recent MEG~II limits~\cite{MEGII:2024urz}, and a definitive confirmation 
or refutation of the $X_{17}$ hypothesis is still lacking.

In this work, we simply comment on the speculative possibility that a 17\,MeV pseudo-scalar resonance 
coupled to electrons, identified with the electrophilic ALP of our framework,  fits the current hints from PADME.
We highlight this with a dotted red line in Fig.~\ref{fig:alp1} and Fig.~\ref{fig:alp3}.
This opens the intriguing possibility that the potential new light boson may be connected to dark matter, acting as a mediator between a dark sector and the SM. Some ideas in this direction were proposed in~\cite{Armando:2023zwz,Darme:2020sjf,AriasAragon2025_LDMStatus} and, for a vector 17 MeV resonance, in~\cite{Alves:2023ree,Barman:2021yaz}. See also~\cite{Alves:2020xhf} for a QCD axion interpretation of the 17 MeV resonance, though not related to dark matter.

Notice that couplings to quarks could also be introduced in our framework to simultaneously address the ATOMKI 
observations~\cite{Krasznahorkay:2015iga,Krasznahorkay:2018ip,Krasznahorkay:2019apb,Krasznahorkay:2019njf,Krasznahorkay:2021joi,Krasznahorkay:2022prc,Krasznahorkay:2023x17,Krasznahorkay:2024universe}. 
While this would make the analysis more involved, it would not substantially affect 
our main conclusions concerning the role of the ALP as a dark matter mediator. Indeed, the thermalization between the ALP and the SM sector would at most benefit from the presence of new interactions, while the CMB and indirect detection limits would be at most comparable to those from the electron-positron  channel. On the other hand, laboratory constraints on the ALP-quark coupling, independent of the role of the ALP as a dark matter mediator, should be considered. Given the speculative nature of this potential resonance, we 
adopt a minimal approach; a more detailed investigation of the $X_{17}$ scenario 
will become meaningful once stronger experimental confirmation is available.
Indeed, a new PADME data-taking campaign with an upgraded detector is planned, aiming to 
significantly improve the sensitivity and determine whether the observed structure is a 
statistical fluctuation or a genuine signal of new physics.

\section{The model with $\theta\neq0$}
\label{sec:theta}

In this section, we consider the impact of a non-vanishing $\theta$ angle in Eq.~\eqref{eq:LagSIMP}. We focus on the case of degenerate fermion masses, while we briefly comment on scenarios with non-degenerate spectra in Appendix~\ref{app:resonant}.
{The $\theta$ angle of the dark sector is not constrained by any experimental data, so that $\theta\sim\mathcal{O}(1)$ is a viable choice. The Chiral Lagrangian in the presence of a sizable $\theta$ has been discussed also in~\cite{Kamada:2017tsq,Garcia-Cely:2025flv}.}

The $\theta$ phase induces new CP-violating interactions among the dark pions and the ALP.
More specifically, we get an extra CP-violating cubic interaction

\begin{equation}\label{eq:thetaLag}
    \frac{m_\pi^2\tan\theta}{4f_a}a{\rm Tr}[\pi^2]=\frac{m_\pi^2\tan\theta}{2f_a}a\pi^a\pi^a\,.
\end{equation}

Notice that, since for $N_f=2$ both ${\rm Tr}[\pi^3]=0$ and ${\rm Tr}[\pi^5]=0$, no $\theta$-induced cubic and quintic  interaction are generated. For a generic choice of gauge group and $N_f$, these interactions are present and may contribute significantly to the DM relic abundance~\cite{Kamada:2017tsq,Garcia-Cely:2024ivo}.

The operator in Eq.~\eqref{eq:thetaLag} provides a novel channel for the thermalization of dark matter. Indeed, the elastic scatterings  $\pi e^\pm\to \pi e^\pm$, mediated by the $t$-channel exchange of the ALP field, lead to a direct  energy transfer from the dark matter sector to the SM, which can be efficient enough to establish kinetic equilibrium. This allows us to overcome the two-step thermalization discussed in the previous section. Furthermore, such thermalization mechanism is no longer sensitive to the number density of the ALP, which only acts as an off-shell mediator, so that a heavy ALP no longer corresponds to an exponentially suppressed rate. As a result, a sizable $\theta$ angle opens the parameter space for $m_a>m_\pi$, which is complementary to the one studied in the previous section. We focus on this regime in the following. Notice that the same coupling in Eq.~\eqref{eq:thetaLag} gives rise to dark matter annihilations into electron-positron pairs, mediated by $s$-wave exchange of the ALP. These are subject to strong constraints, especially close to $m_a\sim 2m_\pi$.

\subsubsection{Thermalization}

The relevant processes for dark matter thermalization are the elastic scatterings of dark matter with electrons and positrons. These occur via the interaction in Eq.~\eqref{eq:thetaLag} and the $t$-channel exchange of the ALP. This is similar to what happens in models where SIMP dark matter interacts with the SM via a dark photon portal~\cite{Hochberg:2015vrg,Katz:2020ywn,Garcia-Cely:2025flv}. The amplitude of the process is 

\begin{equation}
\overline{\left|\mathcal{M}\right|_{t}^{2}}=-\frac{ g_{ae}^{2}g_{\pi\pi a}^{2}t}{\left(m_{a}^{2}-t\right)^{2}}\,,
\end{equation}
where $g_{\pi\pi a}=m_\pi^2\tan(\theta)/f_a$, the bar denotes the average over the initial state and we sum over the final state.  
In the limit of massless leptons and heavy ALP, $m_a\gg m_\pi$, the corresponding cross section is 

\begin{equation}\label{eq:tchannel}
\sigma v(\pi e^\pm\to\pi e^\pm)\simeq \frac{g_{ae}^2m_\pi^2 p_e^2\tan^2\theta}{8\pi m_a^4 f_a^2}\,,
\end{equation}
where $p_e$ is the momentum of the SM fermion. Dark matter particles thermalize with the SM bath as long as~\cite{Hochberg:2015vrg,Katz:2020ywn}

\begin{equation}\label{eq:thermcond}
\frac{5\zeta\left(5\right)}{4}\frac{T_{{}}}{m_{\pi}}\Gamma_{{\rm scatt}}\gtrsim H\,,
\end{equation}
where the scattering rate is given by 

\begin{equation}
    \Gamma_{\rm scatt}=\sum_{f=\{e^+,e^-\}}n_f^{\rm eq}\langle{\sigma v(\pi f\to\pi f)\rangle}\,.
\end{equation}
In the limit of heavy ALP, we get 

\begin{equation}\label{eq:thermGamma}
    \Gamma_{\rm scatt}\overset{    }{\simeq}\frac{45 \zeta(5) g_{ae}^2m_\pi^2T^5\tan^2\theta}{8\pi^3 m_a^4 f_a^2}\,,
\end{equation}
while the general expression is obtained by integrating numerically the cross section. 
We require that thermalization is efficient at least until dark matter freeze-out, computing the condition in Eq.~\eqref{eq:thermcond} for $T=T_{\rm fo}$. Notice that the heavy ALP approximation in Eq.~\eqref{eq:thermGamma} works very well in all the interesting regions of the parameter space since $t\lesssim \mathcal{O}(T_{\rm fo}^2)$ and $m_a\gg T_{\rm fo}\sim m_\pi/20$.

\begin{figure*}[t]
\!\!\!\includegraphics[width=0.45\textwidth]{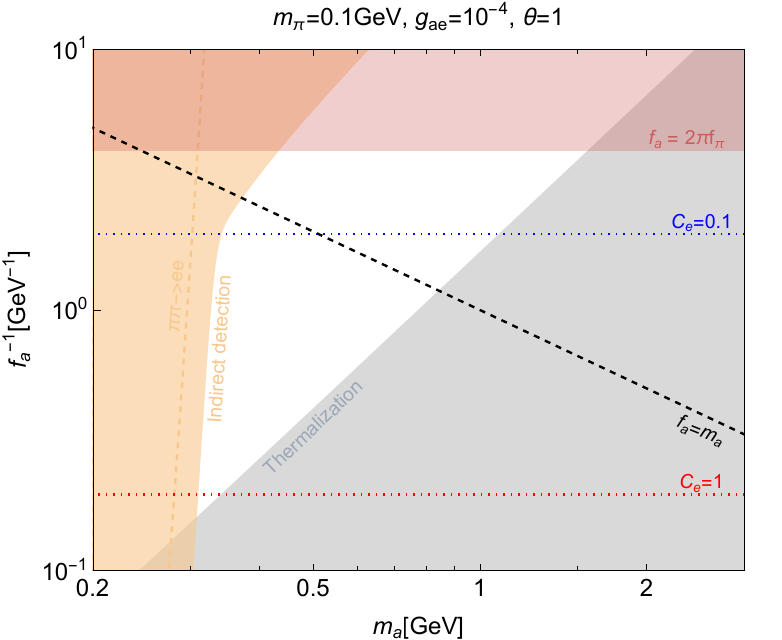}
\includegraphics[width=0.45\textwidth]{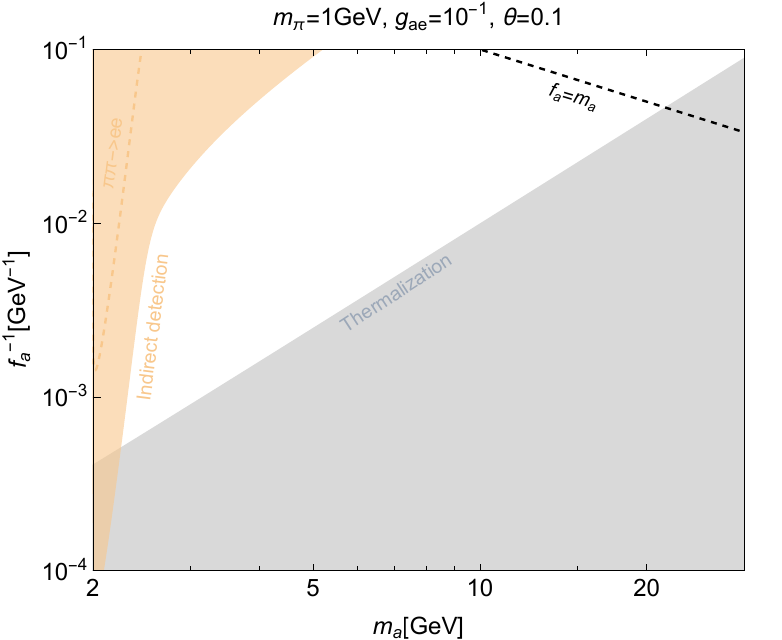}
\,\,\,\,\,\,
\caption{Summary of the constraints in the case of a non vanishing $\theta$ angle and $m_a>m_\pi$. The value of $g_{ae}$ is chosen so that all constraints from ALP laboratory searches are satisfied. In the gray region the dark matter does not thermalize with the SM efficiently, while the orange region is excluded by CMB and indirect detection constraints. 
The available parameter space corresponds to the white region. See the main text for details. } 
\label{fig:alp2}
 \end{figure*}

\subsubsection{Constraints}

The operator in Eq.~\eqref{eq:thetaLag} also provides a new annihilation channel for dark matter particles, namely the process $\pi\pi\to e^+e^-$, which is mediated by the $s$-wave exchange of the ALP. 
The corresponding amplitude is

\begin{equation}\label{eq:ampltis}
\overline{\left|\mathcal{M}\right|_{s}^{2}}=\frac{2g_{ae}^{2}g_{\pi\pi a}^{2}s}{N_\pi\left(s-m_{a}^{2}\right)^{2}}\,.
\end{equation}
In the limit of heavy ALP, this provides the thermally averaged cross section 

\begin{equation}\label{eq:id2}
    \langle{\sigma v\rangle}_{\pi\pi}\equiv\langle{\sigma v(\pi\pi\to e^+e^-)\rangle}\overset{m_a\gg m_\pi}{\simeq}\frac{g_{ae}^2m_\pi^4\tan^2\theta}{4 \pi N_\pi m_a^4f_a^2}\,.
\end{equation}
This approximation fails close to the resonant region $m_a\gtrsim 2m_\pi$, where Eq.~\eqref{eq:ampltis} diverges in the non-relativistic limit $s\simeq 4m_\pi^2$, and the propagator must be ``dressed", adding the extra term $m_a^2\Gamma_a^2$ in the denominator. The thermally averaged cross section is then obtained upon numerical integration of the amplitude. 

In analogy with the previous section, we must ensure that dark‑matter annihilations into SM fermions remain sub‑leading compared to the $3\pi\to2\pi$  annihilations and do not alter the relic abundance. This requirement is satisfied as long as $n_\pi\langle{\sigma v\rangle}_{\pi\pi}\lesssim H|_{T_{\rm fo}}$. 
Furthermore, the CMB and indirect detection constraints discussed earlier apply directly to $\langle{\sigma v\rangle}_{\pi\pi}$. 
It is worth noting that, unlike to scenarios with a dark photon portal, where dark matter annihilations are typically $p$-wave suppressed, in our setup the annihilations proceed via an $s-$wave cross section and provide the strongest constraints on the parameter space.

The elastic scattering of dark matter on electrons, relevant for thermalization, could in principle provide a channel for the direct detection of sub-GeV dark matter. In the case of ALPs in Eq.~\eqref{eq:LagALP}, however, the $\gamma_5$ structure of the electron coupling suppresses the elastic cross section by the dark-matter velocity, $\sigma_{\pi e}\sim v^2$, placing it far below the sensitivity of current direct-detection experiments.
\gl{More precisely, the DM-electron elastic cross section is given by 

\begin{equation}
    \sigma_{e}\simeq\frac{\alpha_{\rm em}^{2}g_{ae}^{2}g_{\pi\pi a}^{2}}{16\pi}\frac{m_e^2}{m_{\pi}^{2}m_{a}^{4}}\,,
\end{equation}
where we used that the typical electron velocity is $v_e\sim\alpha_{\rm em}> v_{\rm DM}\sim 10^{-3}$ and the mediator is heavy, $m_a\gg m_e$. The strongest constraints are reported in Ref.~\cite{XENON:2026qow} and can be re-casted providing lower limits on the scale $f_a$ (as a function of $m_a$ and $m_\pi$). The parameter space of our interest, depicted in Fig.~\ref{fig:alp2}, lies well within the current limits (in particular, for any choice of DM and ALP masses shown in Fig.~\ref{fig:alp2}, $f_a$ is  many order of magnitude larger than the corresponding lower limit).}

{Finally, the interaction in Eq.~\eqref{eq:thetaLag} gives rise to ALP decays to dark matter, $a\to\pi\pi$,  forbidden if $\theta=0$. If $m_a>2m_\pi$ the decay is kinematically open with rate

\begin{equation}\label{eq:inv}
    \Gamma(a\to\pi\pi)=\frac{N_\pi m_\pi^4\tan^2\theta}{32\pi f_a^2m_a}\sqrt{1-\frac{4m_\pi^2}{m_a^2}}\,.
\end{equation}
As long as $\Gamma(a\to 2\pi)\ll\Gamma(a\to e^+e^-)$, which corresponds to $\sqrt{N_\pi}\tan\theta/g_{ae}\ll 2m_af_a/m_\pi^2$, the ALP decays mostly to SM fermions, see the Appendix~\ref{app:decays} for more details. In such a case, the constraints for electrophilic ALPs can be safely applied. In particular, large values of $g_{ae}$, up to the perturbative unitarity limit ($g_{ae}\lesssim\sqrt{8\pi/3}$), are unconstrained for $m_a\gtrsim 250$ MeV.
In the opposite regime, namely for large enough values of  $\theta/g_{ae}$, 
the ALP dominantly decays to DM particles. 
The phenomenology is different, and the strongest constraints for $m_a$ in the 0.2-8 GeV range come from missing-energy searches at BaBar, which set an upper limit $g_{ae}\lesssim10^{-4}$, see Refs.~\cite{Darme:2020sjf,Arias-Aragon:2025qod,AriasAragon2025_LDMStatus}. 

}

We show the available parameter space in Fig.~\ref{fig:alp2}. We chose values of $g_{ae}$ such that all laboratory and astrophysical constraints are safely evaded. 
In the gray region, dark matter scatterings off electrons and positrons via the $t$-channel exchange of the ALP are unable to keep kinetic equilibrium until freeze-out. The orange region is excluded by CMB and indirect detection constraints. Such constraints become much stronger as we get closer to the resonant region $m_a\gtrsim2m_\pi$.
Above the orange dashed line, dark matter annihilation to SM fermions dominates over $3\pi\to2\pi$ annihilation, which occur only in regions which are already excluded. Finally, the effective Lagrangian description is not consistent in the red region, where $f_a<2\pi f_\pi$ (only visible in the plot on the left).

The allowed parameter space corresponds to the white region. The presence of a sizable $\theta$ angle allows the thermalization of the dark sector through a heavy ALP, with $m_a>2m_\pi$ and above the GeV scale, otherwise inaccessible if $\theta=0$. 
Furthermore, values of $f_a$ as low as $\sim$ 500 MeV are allowed in this scenario.

{In the left plot, $m_\pi = 100$ MeV and the ALP decays mostly to DM.
The allowed region of the parameter space is  compatible with the scenario in which the ALP is a pseudo-Goldstone boson. To emphasize that, we show two dotted lines corresponding to $C_e=g_{ae} f_a/m_e=0.1,1$ and a dashed line below which $m_a<f_a$.

As the dark matter mass increases, larger values of $f_a$ and $g_{ae}$ are needed to satisfy the constraints,  so that  $g_{ae}/m_e\gg 1/f_a$. Such large values of  $g_{ae}/m_e$
can be realized if the ALP is not a pseudo-Goldstone boson.
In the right plot we show an example for $m_\pi=1$ GeV. For the chosen values of $g_{ae}$ and $\theta$, the ALP decays mostly to electron-positron pairs.
}

\section{Conclusions}\label{sec:conclusions}

In this work, we have considered the scenario in which dark matter is a dark pion emerging from a confining dark sector, connected to the SM via an ALP mediator. The main novelties of this work are the hypothesis that the ALP is electrophilic, i.e., only couples to electrons at tree level, and the topological $\theta$ angle of the dark confining sector is non-vanishing.

The electrophilic nature of the ALP allows us to relax the experimental constraints based on the axion-photon coupling and enlarges the available parameter space for ALP-mediated SIMP dark matter. In particular, we have shown that an ALP as light as $\mathcal{O}(10)$ MeV can successfully keep the dark matter in kinetic equilibrium with the SM bath while evading the most stringent constraints from CMB measurements, indirect detection and laboratory searches. We have emphasized that an ALP with mass in the range $m_a \sim 10$--$20~\text{MeV}$ {and coupled to electrons} is not only viable but also particularly motivated, as it overlaps with the possible $17~\text{MeV}$ resonance reported by the PADME experiment. 

Finally, we have shown that the inclusion of a $\theta$ angle further enlarges the parameter space, allowing ALPs heavier than the dark matter and above the GeV scale. In such a case, dark matter thermalization proceeds via elastic scatterings off electrons and positrons, directly induced by the $\theta$ parameter itself. In Appendix~\ref{app:resonant}, we briefly comment on the model-building challenges for an ALP-mediator within the dark matter model of Ref.~\cite{Garcia-Cely:2024ivo}, which deserve further investigation.

\section{Acknowledgements}

We would like to thank Luca di Luzio and Robert Ziegler for a careful reading of the manuscript and their useful comments. We also thank Enrico Graziani for his valuable input on the Belle II prospects. Moreover, we thank the referee for his/her valuable comments. The work of FM  is supported by the European Union-Next Generation EU and by the Italian Ministry of University and Research (MUR) via the
PRIN 2022 project No. 2022K4B58X-AxionOrigins.
The work of GL is supported by the INFN Cabibbo Fellowship, call 2025. 
The work of JHG is supported by the Program of China Scholarship Council (Grant No.202506020156).
This article is based upon work from COST Action COSMIC WISPers CA21106, supported by COST (European
Cooperation in Science and Technology).

\vspace{0.2cm}

\appendix

\section{ALP-electron thermalization}\label{app:alpelectron}

In this appendix we derive the condition for an efficient ALP-SM thermalization. We follow closely the analysis in the Appendix C of Ref.~\cite{Hochberg:2018rjs} adopting their notation. We review their derivation of Eq.~\eqref{eq:therm1} and extend their results to the case of scatterings $ae^\pm\to\gamma e^\pm$. We denote the SM temperature as $T$, while the ALP temperature as $T_a$. The equilibrium label denotes quantities evaluated at the SM temperature $T$. 
The relevant Boltzmann equations are 

\begin{widetext}
\begin{align}
\dot{n}_{a}+3Hn_{a}	&=-\bigg(\langle{\Gamma_a\rangle}_{T_a}n_a-\langle{\Gamma_a\rangle}_Tn_a^{\rm eq}\bigg)-2n_e^{\rm eq}\bigg(\left\langle \sigma v_{ae}\right\rangle_{T_a} n_{a}-\left\langle \sigma v_{ae}\right\rangle_{T}n_{a}^{{\rm eq}}\bigg)\\
\dot{\rho}_{a}+3H\left(\rho_{a}+P_{a}\right)	&=-m_a\Gamma_a(n_a-n_a^{\rm eq})-2n_e^{\rm eq}\bigg(\left\langle \sigma v_{ae}E_{a}\right\rangle_{T_a}n_{a}-\left\langle \sigma v_{ae}E_{a}\right\rangle_{T}n_{a}^{{\rm eq}}\bigg)
\,,
\end{align}
\end{widetext}
where we used the fact that $e^\pm$ follow their equilibrium distribution and the factor of 2 in front of $n_e^{\rm eq}$ takes into account the contribution of both electrons and positrons. The thermally averaged decay rate is defined as
\begin{equation}
\left\langle\Gamma_a\right\rangle_T=\frac{m_a}{\langle{E_a\rangle}_T}\Gamma_a=\frac{K_1(m_a/T)}{K_2(m_a/T)}\Gamma_a\,,
\end{equation} while 
\begin{widetext}
\begin{align}
&\langle{\sigma v_{ae}\rangle}_T=\frac{\frac{T}{32\pi^{4}}\int_{(m_a+m_e)^2}^{\infty}ds\sqrt{s}{ K}_{1}\left(\frac{\sqrt{s}}{T}\right)\int{\rm d}t\frac{\overline{\left|\mathcal{M}_{ae}\right|^2}}{8\pi s}}{n_{a}^{{\rm {\rm eq}}}n_{e}^{{\rm {\rm eq}}}}\,,\\
&\left\langle \sigma v_{ae} E_a\right\rangle_T =\frac{\frac{T}{32\pi^{4}}\int_{(m_a+m_e)^2}^{\infty}ds\sqrt{s}{ K}_{1}\left(\frac{\sqrt{s}}{T}\right)\int{\rm d}t\frac{\overline{\left|\mathcal{M}_{ae}\right|^{2}}}{8\pi s}\left(\frac{s+m_{a}^{2}-m_{e}^{2}}{2\sqrt{s}}\right)}{n_{a}^{{\rm {\rm eq}}}n_{e}^{{\rm {\rm eq}}}}\,,
\end{align}
\end{widetext}
are the thermally averaged cross section and the energy transfer rate between the ALP and the SM sector in each scattering process, where the last factor in parenthesis is the ALP c.o.m. energy.

The Boltzmann equations can be written as
\begin{widetext}
\begin{align}
&-T\frac{\partial n_{a}}{\partial T}+3n_{a} =-\frac{m_a\Gamma_a}{H} n_a\left(\langle{E_a^{-1}\rangle}_{T_a}-\langle{E_a^{-1}\rangle}_{T}\frac{n_a^{\rm eq}}{n_a}\right)-\frac{2n_{e}^{{\rm eq}}n_a}{H}\left(\left\langle \sigma v_{ae}\right\rangle_{T_a}-\left\langle \sigma v_{ae}\right\rangle_{T}\frac{n_{a}^{{\rm eq}}}{n_a}\right)\nonumber\\
&\hspace{2.4cm}\equiv-\frac{m_a\Gamma_a}{H}n_ac_n^D-\frac{2n_{e}^{{\rm eq}}n_a}{H}c_{n}^S,
\\
&-T\frac{\partial\left\langle E_{a}\right\rangle_{T_a} n_{a}}{\partial T}+3\left\langle E_{a}\right\rangle_{T_a} n_{a}\left(1+w_{a}\right)  =-\frac{m_a\Gamma_a}{H} n_a\left(1-\frac{n_a^{\rm eq}}{n_a}\right)-\frac{2n_{e}^{{\rm eq}}n_a}{H}\left(\left\langle \sigma v_{ae}E_a\right\rangle_{T_a}-\left\langle \sigma v_{ae} E_a\right\rangle_{T}\frac{n_{a}^{{\rm eq}}}{n_a}\right)\nonumber\\
&\hspace{6.1cm}\equiv-\frac{m_a\Gamma_a}{H}n_ac_\rho^D-\frac{2n_{e}^{{\rm eq}}n_a}{H}c_{\rho}^S\,,
\end{align}
\end{widetext}
where $w_a$ is the ALP equation of state. In the following we focus on very light ALPs, $m_a\ll T_{(a)}$, which are relativistic and satisfy $w_a=1/3$.
    Combining the two equations above and defining $\langle E_a^{-1}\rangle_{T_a}\equiv\alpha(T_a)/\langle{E_a\rangle}_{T_a}$ we get 
\begin{widetext}
\begin{align}
\frac{\partial\langle{E_a}\rangle_{T_a}}{\partial T}=\frac{\langle{E_a\rangle}_{T_a}}{T}+\frac{m_a\Gamma_a}{HT}\left[(1-\alpha)-\frac{n_a^{\rm eq}}{n_a}\left(1-\alpha\frac{\langle{E_a\rangle}_{T_a}}{\langle{E_a\rangle}_{T}}\right)\right]-\frac{2n_e^{\rm eq}}{HT}\left[\langle{E_a\rangle}_{T_a}c_n^S-c_\rho^S\right]\,.
\end{align}
\end{widetext}
The first term in the r.h.s. represents the expansion of the Universe, which tends to cool the ALP bath, while the second term corresponds to the interactions which push towards thermal equilibrium. Thus, in order to guarantee an efficient thermalization, the second term must overcome the first one. We assume that the ALP is close to thermal equilibrium $n_a\simeq n_a^{\rm eq}(T_a)$ and we expand the previous equation for $T_a\simeq T$, so that
\begin{widetext}
\begin{align}
\frac{\partial\langle{E_a}\rangle_{}}{\partial T}=\frac{\langle{E_a\rangle}_{}}{T}-\left[\frac{m_a\Gamma_a}{HT}(2\alpha-3)+\frac{6n_e^{\rm eq}}{HT}\bigg(\langle{\sigma v_{ae}}\rangle\langle{E_a\rangle}_{}-\langle{\sigma v_{ae}E_a\rangle}\bigg)\right]\left(\frac{T_a-T}{T}\right)\,,
    \end{align}
\end{widetext}
where $\langle{\ldots\rangle}=\langle{\ldots\rangle}_{T_a}\simeq \langle{\ldots\rangle}_T$.

\begin{figure*}[t]
\includegraphics[width=0.45\textwidth]{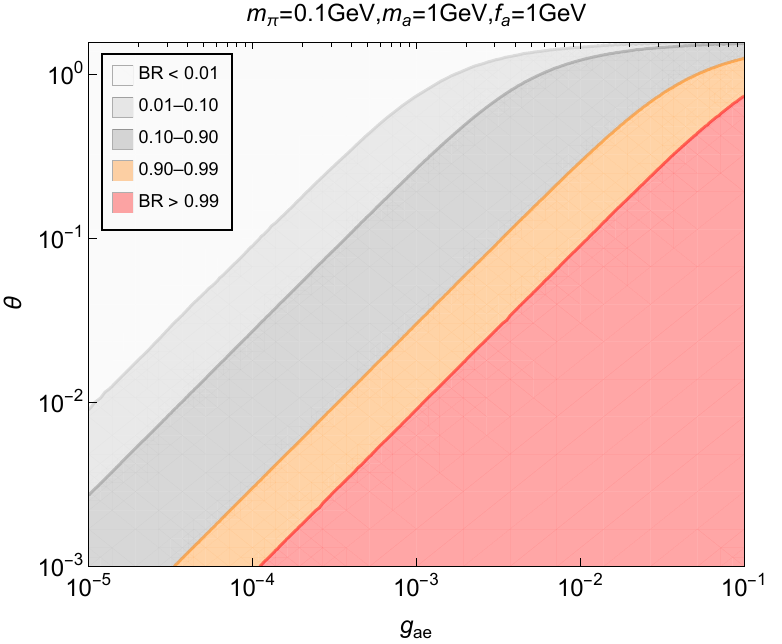} \,\,\,\,\,\,
\includegraphics[width=0.45\textwidth]{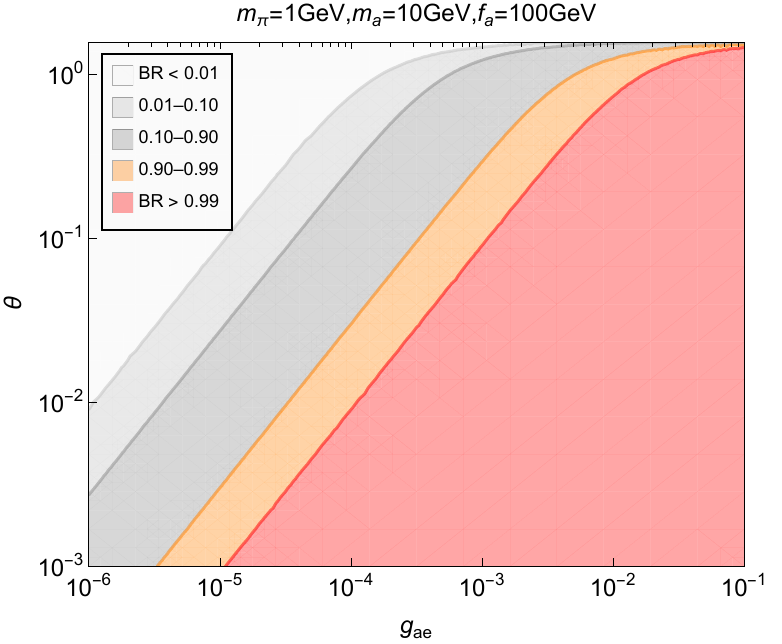}
\caption{The value of branching ratio for the process $a \xrightarrow{} e^+ e^-$ is shown for two representative benchmark points.}
\label{fig:BR}
 \end{figure*}

Decays and inverse processes dominate over scattering if $m_a\gtrsim T_{\rm fo}$, so that thermalization is efficient as long as

\begin{equation}
    \frac{m_a\Gamma_a}{HT}\gtrsim \frac{\langle{E_a\rangle}}{(2\alpha-3)T}\simeq 4\,,
\end{equation}
where we used that for a relativistic particle $\alpha=\pi^6/360\zeta(3)^2$. We recover the result of~\cite{Hochberg:2018rjs}, which can be generalized to the case of heavier ALPs as in Eq.~\eqref{eq:therm1}.

For lighter ALPs, $m_a\ll T_{\rm fo}$, the decay rate is suppressed by the boost factor $m_a/T$, while scatterings are more efficient. Since the ALP is very light in this regime we can always rely on the relativistic approximation, so that thermalization is efficient as long as

\begin{equation}
    \frac{n_e^{\rm eq}}{HT}\bigg(\langle{\sigma v_{ae}}\rangle\langle{E_a\rangle}_{}-\langle{\sigma v_{ae}E_a\rangle}\bigg)\gtrsim\frac{\langle{E_a\rangle}}{6T}\simeq 0.45\,,
\end{equation}
which generalizes Eq.~\eqref{eq:therm2}.

\gl{Notice that, we have assumed that ALPs are described by an equilibrium distribution with a temperature $T_a$, despite the fact that ALPs self-scatterings are quite suppressed with respect to all the other relevant processes. This approach is still justified as long as the interactions of the ALPs with the SM bath are 
strong enough to keep them always very close to thermal equilibrium, as required in our setup.
Furthermore, even if the ALP distribution deviates from the equilibrium shape, there are no velocity-dependent processes at the relevant energies that would significantly influence the heat transfer between the sectors or the evolution of the dark pion number density.
}

\section{ALP decay modes}\label{app:decays}

As long as $m_a<2m_\pi$, the ALP can only decay to visible states, $a\to e^+e^-$, with the decay rate given in Eq.~\eqref{eq:decayrate}. Moreover, if $\theta = 0$, even if $m_a > 2 m_\pi$, the ALP couples to the
dark pions only quadratically, so that the decay into two pions remains
forbidden.

The situation changes once $\theta \neq 0$ and $m_a > 2 m_\pi$. In this
case, the $\theta$ angle induces the interaction in
Eq.~\eqref{eq:thetaLag}, which is linear in the ALP field, and the decay
$a \to \pi\pi$ becomes kinematically allowed.

The decay rate to dark pions is given in Eq.~\eqref{eq:inv}.
The ALP decays mostly to visible states as long as
\begin{equation}
    \frac{\tan\theta}{g_{ae}}\ll \frac{2}{\sqrt{N_\pi}}\frac{m_a}{m_\pi}\frac{f_a}{m_\pi} \left(\frac{1-4 m_e^2/m_a^2}{1-4 m_\pi^2/m_a^2}\right)^{\frac{1}{4}}\,.
\end{equation}
In Fig.~\ref{fig:BR}, we show contours of the branching ratio $\text{BR}(a\to e^+e^-)$ as a function of $g_{ae}$ and $\theta$ for two representative benchmark points.

\section{A comment on the model of Ref.~\cite{Garcia-Cely:2024ivo}}\label{app:resonant}

As a final remark, we briefly comment on  the scenario of non-degenerate quark masses. It has been shown recently that a $\SU(N_c)$ gauge sector with 3 light Dirac quarks of different masses ($m_1\neq m_2\neq m_3$) and a non-vanishing $\theta$ angle can give rise to a meson spectrum with a resonance~\cite{Garcia-Cely:2024ivo,Garcia-Cely:2025flv}. More precisely, the spectrum of the pseudo-Goldstone mesons resembles the one of ordinary QCD. The lightest state, dubbed $\pi^0$, constitutes the dark matter candidate of the model, while the heaviest state, dubbed $\eta$, satisfies $m_\eta=(2+\epsilon)m_{\pi^0}$ with $\epsilon\ll1$. This setup is particularly interesting since the presence of the $\eta$ resonance and the $\theta$ angle  can  simultaneously explain the dark matter relic abundance and provide  velocity-dependent dark matter self-interactions in astrophysical halos. Most precisely, dark matter self-interactions are characterized by a sharp velocity dependence, which predicts large self-interacting cross sections in small astrophysical halos, while evading the stringent constraints derived from observational data from galaxy clusters~\cite{Harvey:2015hha,Bondarenko:2017rfu,Harvey:2018uwf,Sagunski:2020spe,DES:2023bzs}. 
Ref.~\cite{Garcia-Cely:2025flv} showed that, within this ``resonant model", the presence of a dark photon portal successfully establishes thermal equilibrium between the dark matter and the SM bath in the early Universe, ensures the dark matter stability and evades indirect detection limits. An interesting question is whether the ALP portal is also a viable option. 

Here we highlight the main challenges for model-building: first, non-degenerate quark masses unavoidably predict a non-vanishing mixing between the dark matter particle and the ALP (no symmetry prevents it, unlike the model of the main text, which is protected by the global $\Sp(2N_f)$). This can lead to dark matter decays (to $e^+e^-$ if the ALP is electrophilic or to photons otherwise) which may be too fast to ensure cosmological stability of dark matter, unless some cancellation in the mixing angle is provided. Furthermore, a non-vanishing mixing between the ALP and the $\eta$ resonance is also present. This may lead to a resonant production of SM particles via $\pi\pi \to (\eta) \to e^+e^-$ in astrophysical halos, where the self-scattering $\pi\pi \to (\eta)\to \pi\pi$, mediated by an on-shell $\eta$ meson, is resonantly enhanced. This is severely constrained by indirect detection limits which require a small $a$-$\eta$ mixing. Finally, if the ALP is lighter than the dark matter, the $a$-$\eta$ mixing also predicts $\pi\pi\to \pi a$ annihilations, which are enhanced with respect to $\pi\pi\to a a$ by a factor of $(f_a/f_\pi)^2$. As a result, the subsequent ALP decays lead to  
indirect detection limits stronger than the ones depicted in Fig.~\ref{fig:alp1}.

To summarize, the ALP portal faces some potential issues which require dedicated model-building strategies. One interesting possibility is to consider a variant of the model of~\cite{Garcia-Cely:2024ivo}, with a degenerate quark spectrum and a $\eta'$-like meson playing the role of the resonance. In such a setup, dark matter would be absolutely stable, while indirect detection constraints from $\eta'$-$a$ mixing would represent the main challenge to be addressed.

\begin{small}
\bibliography{bibliography}
\end{small}

\end{document}